# Optical Stabilization of Fluctuating High Temperature Ferromagnetism in YTiO$_3$

A.S. Disa[1], J. Curtis[2], M. Fechner[1], A. Liu[1], A. von Hoegen[1], M. Först[1], T.F. Nova[1], P. Narang[2], A. Maljuk[3], A.V. Boris[4], B. Keimer[4], A. Cavalleri[1,5]

[1]*Max Planck Institute for the Structure and Dynamics of Matter, Hamburg, Germany*
[2]*John A. Paulson School of Engineering and Applied Sciences, Harvard University, Cambridge, USA*
[3]*Leibniz Institute for Solid State and Materials Research Dresden, Germany*
[4]*Max Planck Institute for Solid State Research, Stuttgart, Germany*
[5]*Clarendon Laboratory, Department of Physics, Oxford University, Oxford, UK*

**In quantum materials, degeneracies and frustrated interactions can have a profound impact on the emergence of long-range order, often driving strong fluctuations that suppress functionally relevant electronic or magnetic phases. Engineering the atomic structure in the bulk or at heterointerfaces has been an important research strategy to lift these degeneracies, but these equilibrium methods are limited by thermodynamic, elastic, and chemical constraints. Here, we show that all-optical, mode-selective manipulation of the crystal lattice can be used to enhance and stabilize high-temperature ferromagnetism in YTiO$_3$, a material that exhibits only partial orbital polarization, an unsaturated low-temperature magnetic moment, and a suppressed Curie temperature, $T_c$ = 27 K. The enhancement is largest when exciting a 9 THz oxygen rotation mode, for which complete magnetic saturation is achieved at low temperatures and transient ferromagnetism is realized up to $T_{neq}$ > 80 K, nearly three times the thermodynamic transition temperature. First-principles and model calculations of the nonlinear phonon-orbital-spin coupling reveal that these effects originate from dynamical changes to the orbital polarization and the makeup of the lowest quasi-degenerate Ti $t_{2g}$ levels. Notably, light-induced high temperature ferromagnetism in YTiO$_3$ is found to be metastable over many nanoseconds, underscoring the ability to dynamically engineer practically useful non-equilibrium functionalities.**

The macroscopic properties of quantum materials are determined by a delicate tension between microscopic elements, the most relevant being the crystal structure, the magnetic state of the constituent electrons, and the orbitals that they occupy. Degeneracy and its lifting play a particularly fundamental role. For instance, Jahn-Teller distortions lift orbital degeneracies in certain correlated insulators and lead to the stabilization of long-ranged spin and orbital [1, 2]. In cases where degeneracies are not effectively lifted, long-range electronic order is often depressed, and precursor fluctuations are observed far above the thermodynamic transition temperature [3, 4]. In particular, recent work has highlighted the key role of the orbital configuration in determining the stability of superconducting, magnetic, and other electronically ordered phases of correlated materials [5-9].

One of the best examples of this behavior is found in the family of Mott insulating rare-earth titanates ($R$TiO$_3$) [10]. The low-energy physics of this system is dictated by a single Ti electron occupying a manifold of $d$ orbitals with $t_{2g}$ symmetry, namely the $d_{xz}$, $d_{yz}$, and $d_{xy}$ orbitals (Fig. 1a). Within the Goodenough-Kanamori-Anderson picture, the superexchange process is expected to favor antiferromagnetic (AF) or ferromagnetic (FM) spin interactions for electrons hopping between the same or orthogonal orbitals, respectively, leading to a strong dependence on the details of the bonding and crystal field environment at each Ti site. The ideal cubic perovskite structure with degenerate $t_{2g}$ levels would create a highly frustrated ground state with large composite spin-orbital fluctuations [11, 12]. In reality, Jahn-Teller and GdFeO$_3$-type orthorhombic structural distortions lift the $t_{2g}$ degeneracy in the titanates, pushing the system towards a static orbital ordering pattern and tipping the balance in favor of a particular magnetic state [13-18].

Due to this intricate balance of interactions, magnetism in the titanates is highly susceptible to small changes in crystal structure. For example, upon decreasing the rare-earth ion size from $R$ = La to $R$ = Y (which increases the magnitude of structural distortions), one observes a crossover from an AF to an FM ordered phase at low temperatures [19]. Moreover, numerous experimental studies show evidence for magnetic instabilities over a wide temperature range, which may be tied to fluctuations of the lattice and/or orbitals [20-24].

In ferromagnetic YTiO$_3$ ($T_c \approx$ 27K), such fluctuations manifest in several ways. On the one hand, the magnetic moment at low temperatures is found to saturate well below the ideal spin-½ limit, even for $T \ll T_c$ (Fig. 1b) [25]. On the other hand, magnetic contributions to the specific heat and thermal expansion are observed up to more than 3 × $T_c$ [26]. In addition, anomalous phonon-frequency shifts attributable to spin correlations have been detected at similarly high temperatures [27].

The presence of these experimental signatures suggests that under equilibrium conditions, long-range ferromagnetic ordering in YTiO$_3$ may be stifled by competing interactions or dynamical fluctuations. Such effects effectively limit the utility of this and related correlated quantum materials for building novel technologies. Thus, a question of key importance is whether, and by what means, it is possible to harness their harmful fluctuations to attain enhanced functional properties.

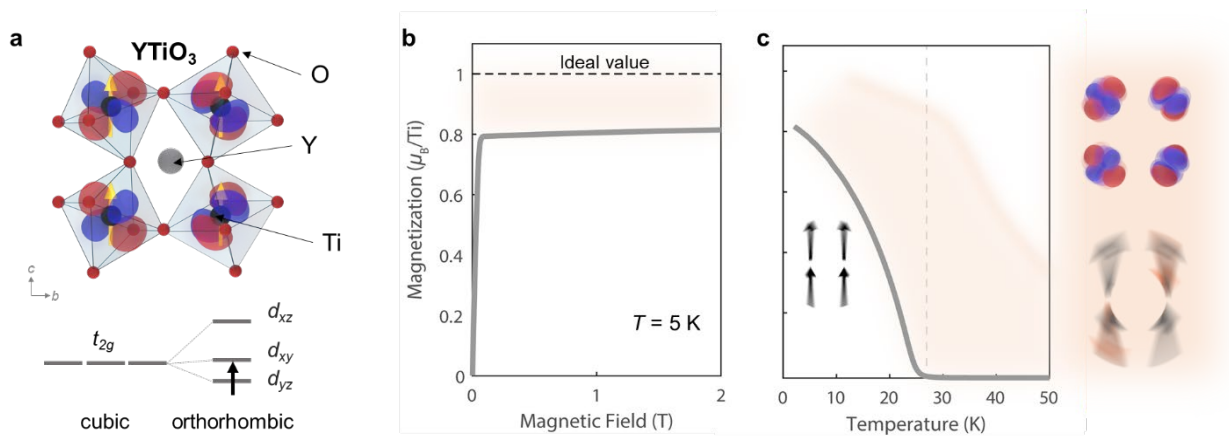

*Figure 1. Fluctuating spin-orbital order in YTiO$_3$. (a) Crystal structure along with the associated low-temperature ferromagnetic and orbital ordering pattern. The orthorhombic structure determines the crystal field splitting and orbital mixing of the Ti $t_{2g}$ levels on each Ti site (b) Magnetization as a function of magnetic field measured at $T \ll T_c$, which saturates at high fields to ~0.8 µ$_B$/Ti – well below the theoretical limit. Fluctuations of the lattice and orbitals weaken ferromagnetic order through competing antiferromagnetic interactions, manifesting as a diminished magnetic moment and reduced critical temperature. (c) Magnetization as a function of temperature. Spin correlations extend well above $T_c$ = 27 K. The inset schematically shows the fluctuating orbital and spin configurations within the shaded region above $T_c$.*

In this paper, we take advantage of the strong spin-lattice coupling of $YTiO_3$ to address this problem, demonstrating the ability to enhance ferromagnetism with light. In particular, we resonantly excite vibrational modes of the lattice using intense terahertz frequency optical pulses. Deformations of the crystal structure not found in equilibrium become possible through the light-matter interaction, which can be engineered by selectively exciting specific phonons [28-30]. Previously, this technique has proven to be an effective tool to alter both local electronic states and their interactions in correlated materials [31-35], and we exploit it here to control orbital/magnetic order in $YTiO_3$ through the lattice [36, 37].

We restrict ourselves to *b*-axis modes with $B_{2u}$ symmetry (see Fig. 2a), which were estimated to have the strongest spin coupling from linear infrared spectroscopy. The coupling strength varies both in sign and magnitude across the 9 $B_{2u}$ modes (see Supplementary Material). For our experimental study, we focused on the modes with center frequencies near 4, 9, and 17 THz, which are relatively well separated from other modes, have comparable oscillator strengths, and are predicted to display significantly different responses [38]. The atomic motions associated with these modes are shown in Fig. 2b.

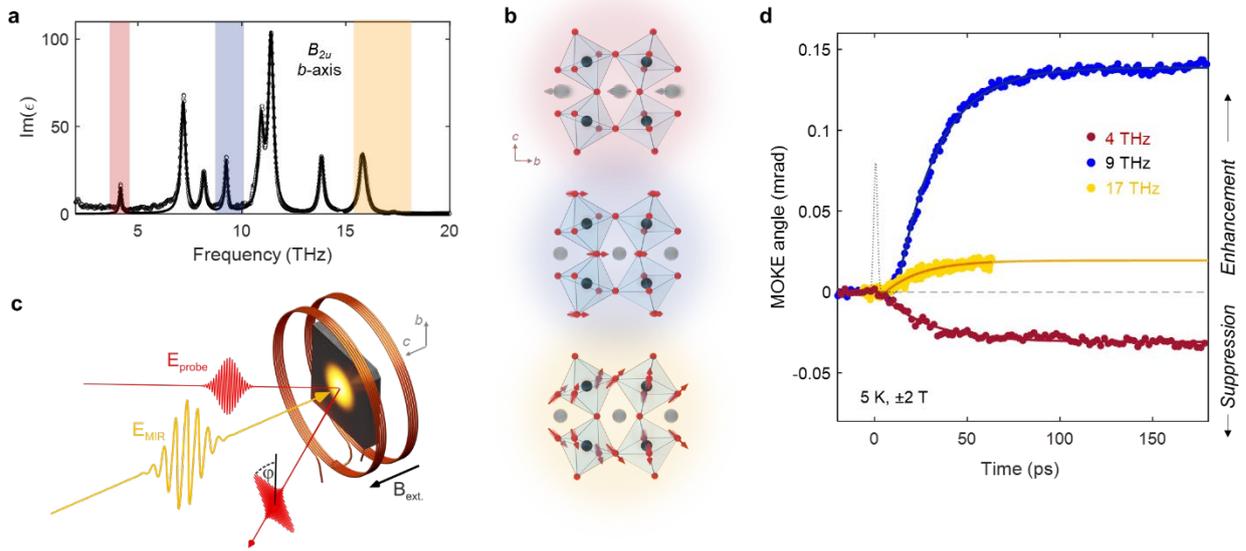

Figure 2. Phonon-selective control of ferromagnetism in $YTiO_3$. (a) Infrared optical spectrum of b-axis vibrational modes (black). The three phonons pumped in this experiment are shaded in red (4 THz), blue (9 THz), and yellow (17 THz). (b) The eigendisplacements corresponding to the pumped vibrational modes in (a): the low-frequency mode primarily involves anti-polar motions of the Y ions, while the two higher frequency modes mainly consist of displacements of the apical and equatorial oxygens, respectively, within the $TiO_6$ octahedral cage. (c) Depiction of the experimental set up for the time-resolved MOKE measurements. (d) Pump-induced changes of the MOKE angle ($\varphi_M = \frac{\varphi(+H) - \varphi(-H)}{2}$) for the three different phonon excitations. The black dotted line shows the pump pulse envelope.

The excitation pulses were generated using a recently developed terahertz source based on chirped pulse difference-frequency generation in an organic crystal, producing narrow bandwidth and high intensity pulses over the entire frequency range where phonon resonances are found, spanning far to mid-infrared wavelengths [39]. The pulse durations range from ∼150 fs for the highest frequency excitation to ∼350 fs for the lowest, and the pump fluence was kept at 5 mJ/cm² (corresponding to peak electric fields of ∼2-4 MV/cm) for all measurements, unless otherwise noted [40]. For these field strengths, we estimate the oscillatory displacements for these modes to range between 5 and 10 pm, far above the rms thermal fluctuations of the equilibrium state (see Supplementary Material).

To determine the changes to the magnetic state of $YTiO_3$ induced by phonon excitation, we carried out time-resolved magneto-optic Kerr effect (MOKE) measurements. A schematic of the experimental pump-probe setup is shown in Fig. 2c. The terahertz excitation pulses were focused at normal incidence to the (001) surface of a YTiO$_3$ single crystal, propagating parallel to the ferromagnetic c axis, and linearly polarized along the b axis to excite the relevant $B_{2u}$ modes. The MOKE signal was determined from the polarization rotation of a time-delayed probe pulse reflected from the sample as a function of external magnetic field ($H \parallel c$). In this geometry, the MOKE angle, $\varphi_M$ (the component of the polarization rotation antisymmetric with respect $H$), is proportional to the c-axis magnetization.

Figure 2d shows the pump-induced change in the MOKE angle, $\Delta\varphi_M$, as a function of time for each of the three excitation frequencies, taken at $T$ = 5 K with $H$ = ±2 T. The signal was calibrated such that positive $\Delta\varphi_M$ signifies an increase in the already existing ferromagnetic magnetization in equilibrium, while negative $\Delta\varphi_M$ signifies a reduction. For all cases, we observed that the pump initiated a gradual change in the magnetization that plateaued at a maximum value on a time scale of roughly 50 ps, after which it remained stable with no measurable decay through the duration of our measurement window (200 ps). We estimate from fitting data taken with longer time windows that the lifetime of this saturated state is at least several nanoseconds (see Supplementary Material).

Notably, the sign and strength of the effect was found to differ significantly depending on the phonon being pumped. For the 4 THz pump, $\Delta\varphi_M$ was negative and relatively small (plateauing at -0.04 mrad), indicating that ferromagnetic order was weakened by the phonon excitation. On the other hand, the positive signal for the 9 and 17 THz modes implies that driving these phonons enhanced the ferromagnetism of YTiO$_3$. In addition, the 9 THz phonon was three times as effective in producing a change in magnetization as the other phonons. These results point to the existence of a long-lasting non-equilibrium state whose magnetic properties are highly sensitive to the structural distortions induced by the resonant THz drive.

The magnetic field dependence of the time-resolved MOKE signal provides further insight into this phenomenon, as illustrated in a series of measurements examining the case of 9 THz phonon excitation. The time evolution of $\Delta\varphi_M$ is similar for all magnetic field strengths, but the maximum value attained at long time delays varies strongly (Fig. 3a). We can extract the *M-H* behavior in the plateau region (*t* >100 ps) by converting $\Delta\varphi_M$ into a magnetization using the calibration procedure outlined in the Supplementary Material. Figure 3b shows that, compared to equilibrium YTiO$_3$, the saturation magnetization at high magnetic fields (> ~0.2 T) is roughly 20% greater in the non-equilibrium state, nearly reaching 1 $\mu_B$/Ti. Overall, the light-driven *M-H* dependence resembles that of ideal ferromagnetic YTiO$_3$ absent competing interactions, hinting at the possibility that the phonon-mediated enhancement stems from a reduction in the fluctuations purported to suppress the ferromagnetic order in equilibrium.

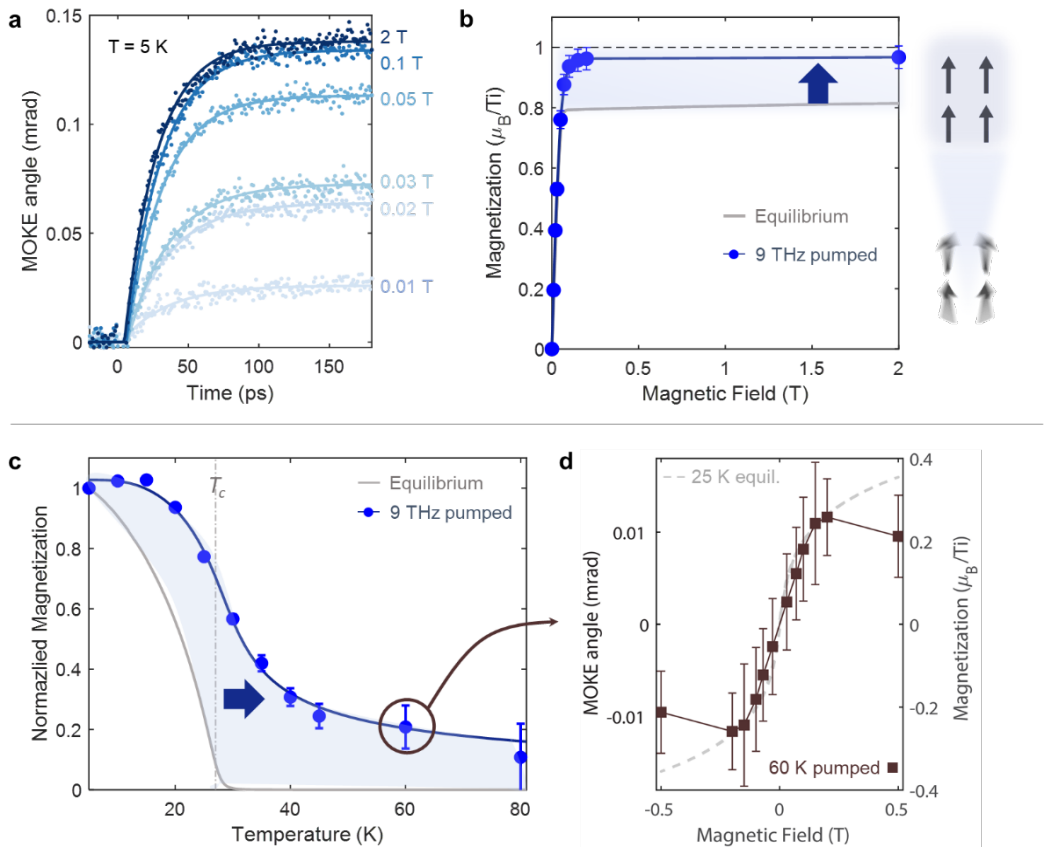

*Figure 3. Characterization of the 9 THz pumped non-equilibrium state. (a) Time-resolved MOKE signal for different magnetic fields at T = 5 K. (b) Extracted sat non-equilibrium magnetization (at t > 100 ps) (blue circles), compared to the equilibrium magnetization (gray solid line). The light-enhanced magnetic state saturates to nearly the ideal spin-½ limit, suggesting a suppression of spin fluctuations (inset). (c) Temperature dependence of the non-equilibrium magnetization (blue circles). The light-induced effect extends up to at least 80 K, well above the equilibrium case (gray solid line). (d) Saturated pump-induced MOKE signal as a function of magnetic field at T = 60 K. The field dependence follows that of equilibrium YTiO$_3$ below $T_c$, indicating the stabilization of a high-temperature ferromagnetic state.*

Having observed the enhancement in ferromagnetism well below the equilibrium Curie temperature, we investigated how this dynamical effect evolves as a function of temperature (Fig. 3c). Unlike for the unperturbed system, where the magnetization drops to zero at $T_c$, we found that pumping YTiO$_3$ at 9 THz induced a magnetization up to temperatures in excess of 80 K – nearly three times $T_c$ – matching the temperature scale associated with anomalous magnetic correlations found in equilibrium YTiO$_3$ (illustrated as orange region in Fig. 1c) [26]. The magnetic field dependence of the pump-induced signal above $T_c$ reveals a nonlinear *M-H* characteristic reminiscent to that at low temperatures, but with a weaker spin stiffness (Fig. 3d). Hence, the phonon excitation appears to create a transient, non-equilibrium ferromagnetic state above $T_c$ that persists to an effective onset temperature much higher than in equilibrium.

As might be expected from the low temperature behavior, the non-equilibrium onset temperature $T_{neq}$ depends on the phonon being pumped (Fig. 4a). The 17 THz phonon, which also showed an enhancement at low temperatures, produced a $T_{neq}$ larger than $T_c$ (but smaller than for 9 THz), while the 4 THz phonon, for which the low temperature magnetic state was diminished, led to a state with $T_{neq}$ slightly less than $T_c$. The shift in the magnetic onset temperature relative to $T_c$ roughly scales with the change in magnetization found at $T<<T_c$ (see Fig. 2d).

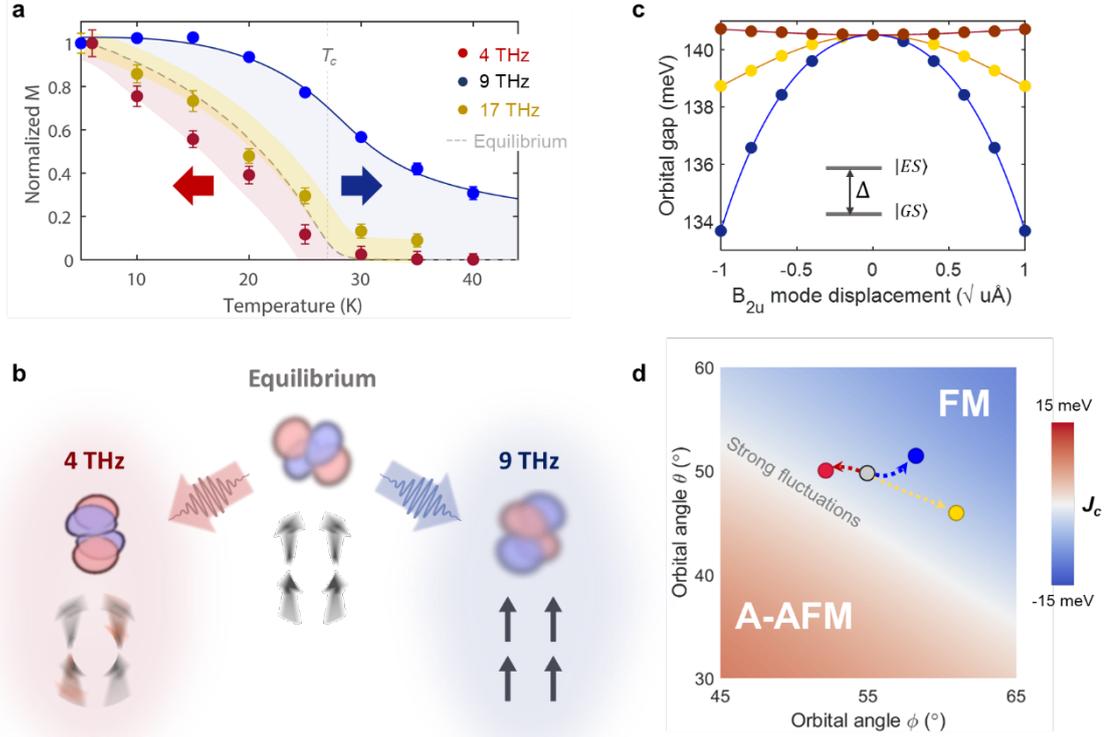

*Figure 4. Possible origin of phonon-driven enhancement of ferromagnetism. (a) Temperature dependence of non-equilibrium magnetization for each pump excitation frequency (colored circles). (b) Illustration of the equilibrium orbital ground state $|GS\rangle$ and the changes ($|GS\rangle_{\text{pumped}} - |GS\rangle_{\text{equil.}}$) induced by driving the 4 THz and 9 THz phonons. The structural and orbital changes push the system farther or closer to the phase boundary, respectively, thereby suppressing or enhancing detrimental spin fluctuations. (c) The energy gap $\Delta$ between $|GS\rangle$ and first excited orbital state $|ES\rangle$ for each of the three excited phonons. The experimental mode displacements are on the order of $1 \sqrt{u}$Å in magnitude. (d) Magnetic phase diagram for varying $|GS\rangle$ of the $Ti_1$ site, where $|GS\rangle = \sin\theta\cos\phi |yz\rangle + \sin\theta\sin\phi |xz\rangle + \cos\theta |xy\rangle$. The calculated equilibrium Case is shown as a grey dot and the corresponding to the distorted structures upon phonon excitation are shown as colored circles. $J_c$ is the average out-of-plane exchange energy.*

Based on the stark differences in the pump-induced response when driving different vibrational modes, the associated structural distortions must be important contributors to the observed dynamical behavior. The most direct mechanism by which driven phonons can influence magnetism is through modifications to the exchange interaction (*J*) caused by the associated atomic motions. We used density functional theory (DFT) to calculate the direct spin-phonon coupling strength for each mode, $\lambda_{spin-ph} = \frac{1}{2}\frac{\partial^2 J}{\partial^2 Q}$, where $Q$ is the phonon amplitude (see Supplementary Material). While the relative values for $\lambda_{spin-ph}$ correlate with the relative signs and magnitudes of the pump-induced effects, the absolute values predict the opposite behavior from what is measured in our experiments. Hence, we rule out a direct spin-phonon coupling mechanism of the form described above and instead consider the fact that the rare-earth titanates exhibit strong spin-orbital-lattice coupling. To model the relevant physics, we developed a Kugel-Khomskii type Hamiltonian (based on Refs. [12, 14, 41]) that takes into account the interaction between these three degrees of freedom within the Ti $t_{2g}$ manifold of YTiO$_3$ (see Supplementary Material). With input from DFT, this model allows us to independently explore the magnetic and orbital landscapes associated with the driven lattice vibrations in a realistic setting.

From these calculations, we ascertain two major contributing factors: the composition of the orbital ground state and the orbital gap (Figs. 4b,c). The orbital ground state of the system (*i.e.* the most occupied $t_{2g}$ orbital admixture) is influenced by the crystal field splitting as well as the intra-site hybridization between orbitals, both of which depend heavily on displacements of the Y and O ions. As discussed in the introduction, the orbital composition dictates the allowed superexchange pathways and, hence, the relative energies of the various types of magnetic order (the two lowest energy being FM and A-type AF). In Fig. 4b, we construct a magnetic phase diagram as a function of the orbital state of the Ti$_1$ site, written in terms of two orbital angles: $\psi_{GS}(\theta, \phi) = \sin\theta \cos\phi \, |yz\rangle + \sin\theta \sin\phi \, |xz\rangle + \cos\theta \, |xy\rangle$. In equilibrium, the orbital ground state lies close to the phase boundary between FM and A-type AF order, contributing to the magnetic fluctuations that suppress FM order in YTiO$_3$. The distorted structures arising from phonon excitation push the orbital ground state either closer to or farther from the phase boundary for the 4 THz mode and 9/17 THz modes, respectively, in agreement with the experimental observations (Fig. 4b). Correspondingly, we find that the orbital energy gap is modulated by the phonons in such a way as to increase or reduce the degree of orbital polarization, respectively, and as a result control the relative importance of the FM favoring superexchange pathways (Fig. 4c). Crucially, these effects are both quadratic in the amplitude of the optically driven phonons and convert oscillatory atomic displacements with zero net distortion into an average *rectified* change of the orbital properties.

Thus, the picture of the non-equilibrium state that emerges is one in which driven lattice distortions nonlinearly alter the orbital ground state and orbital polarization, which in turn modifies ferromagnetism in $YTiO_3$ by amplifying or eliminating phase competition/fluctuations. Although we find excellent qualitative agreement between our theory and experiment, given the apparent importance of fluctuations in the explanation of photo-induced magnetism, extending our model beyond mean field could serve to provide a more quantitatively accurate depiction of the experimental situation. What remains to be understood are the dynamics of the non-equilibrium state and its metastability. Plausibly, the photo-induced ferromagnetic state exhibits domain structure and large spatial fluctuations, especially at high temperatures. The buildup of long-ranged spatial correlations over the probed region may limit the time scale associated with the light-driven magnetic order [42]. Alternatively, the transfer of angular momentum between the lattice and the spin system often acts as a bottleneck for ultrafast magnetic processes, especially in insulators like $YTiO_3$ [43]. In either case, the existence of many competing ordered phases in $YTiO_3$ can provide local free energy minima that trap the driven system into a long-lasting non-equilibrium state [44, 45]. We also note that, as has been previously posited [36, 37], our results point to the possibility of realizing a metastable, non-equilibrium A-type AF phase under an appropriate phonon drive, which has no counterpart in the equilibrium $R$TiO$_3$ phase diagram.

The experiments reported here demonstrate the power of resonant lattice excitation to control electronically ordered phases, by exploiting their strong coupling to crystal structure [31-35, 46]. This result is relevant not just to magnetism in titanates, but to the wide group of quantum materials which display unwanted degeneracies and fluctuating orders. For example, many Cu- and Fe-based superconductors show signatures of remnant correlations at high temperatures indicating that an electronically ordered phase is suppressed in equilibrium, which may be tied to superconductivity [47-50]. The non-equilibrium pathway we demonstrate to enhance an underlying order allows otherwise latent quantum phases to emerge at higher temperatures, bringing their unique functionalities into a technologically useful regime. Finally, the experiments reported here represent a rare case of light-driven symmetry breaking and can be viewed in the same broad class of discoveries as optically enhanced ferroelectricity, superconductivity or charge orders [51-53].

# References

1. K.I. Kugel and D.I. Khomskiĭ, *The Jahn-Teller effect and magnetism: transition metal compounds.* Soviet Physics Uspekhi, 1982. **25**(4): pp. 231-256.
2. Y. Tokura and N. Nagaosa, *Orbital Physics in Transition-Metal Oxides.* Science, 2000. **288**(5465): pp. 462-468.
3. B. Keimer and J.E. Moore, *The physics of quantum materials.* Nature Physics, 2017. **13**(11): pp. 1045-1055.
4. L.F. Feiner, A.M. Oleś, and J. Zaanen, *Quantum Melting of Magnetic Order due to Orbital Fluctuations.* Physical Review Letters, 1997. **78**(14): pp. 2799-2802.
5. L. Feiner and A. Oleś, *Electronic origin of magnetic and orbital ordering in insulating LaMnO$_3$.* Physical Review B, 1999. **59**(5): pp. 3295-3298.
6. N.B. Aetukuri, et al., *Control of the metal–insulator transition in vanadium dioxide by modifying orbital occupancy.* Nature Physics, 2013. **9**(10): pp. 661-666.
7. D. Li, et al., *Superconductivity in an infinite-layer nickelate.* Nature, 2019. **572**(7771): pp. 624-627.
8. A.S. Disa, et al., *Orbital Engineering in Symmetry-Breaking Polar Heterostructures.* Physical Review Letters, 2015. **114**(2).
9. M. Hepting, et al., *Electronic structure of the parent compound of superconducting infinite-layer nickelates.* Nature Materials, 2020. **19**(4): pp. 381-385.
10. M. Mochizuki and M. Imada, *Orbital physics in the perovskite Ti oxides.* New Journal of Physics, 2004. **6**: pp. 154-154.
11. A.M. Oleś, et al., *Spin-Orbital Entanglement and Violation of the Goodenough-Kanamori Rules.* Physical Review Letters, 2006. **96**(14).
12. G. Khaliullin and S. Okamoto, *Theory of orbital state and spin interactions in ferromagnetic titanates.* Physical Review B, 2003. **68**(20).
13. E. Pavarini, et al., *Mott Transition and Suppression of Orbital Fluctuations in Orthorhombic 3d$^1$ Perovskites.* Physical Review Letters, 2004. **92**(17).
14. X.-J. Zhang, E. Koch, and E. Pavarini, *Origin of orbital ordering in YTiO$_3$ and LaTiO$_3$.* Physical Review B, 2020. **102**(3).
15. H. Nakao, et al., *Quantitative determination of the atomic scattering tensor in orbitally ordered YTiO$_3$ by using a resonant x-ray scattering technique.* Physical Review B, 2002. **66**(18).
16. A.C. Komarek, et al., *Magnetoelastic coupling in RTiO$_3$ (R=La,Nd,Sm,Gd,Y) investigated with diffraction techniques and thermal expansion measurements.* Physical Review B, 2007. **75**(22).
17. F. Iga, et al., *Determination of the Orbital Polarization in YTiO$_3$ by Using Soft X-Ray Linear Dichroism.* Physical Review Letters, 2004. **93**(25).
18. J. Akimitsu, et al., *Direct Observation of Orbital Ordering in YTiO$_3$ by Means of the Polarized Neutron Diffraction Technique.* Journal of the Physical Society of Japan, 2001. **70**(12): pp. 3475-3478.
19. T. Katsufuji, Y. Taguchi, and Y. Tokura, *Transport and magnetic properties of a Mott-Hubbard system whose bandwidth and band filling are both controllable: $R_{1-x}Ca_xTiO_{3+y/2}$.* Physical Review B, 1997. **56**(16): pp. 10145-10153.
20. J.G. Cheng, et al., *Transition from Orbital Liquid to Jahn-Teller Insulator in Orthorhombic Perovskites RTiO$_3$.* Physical Review Letters, 2008. **101**(8).

21. B. Li, et al., *Dynamic Distortions in the YTiO3 Ferromagnet.* Journal of the Physical Society of Japan, 2014. **83**(8).
22. N.N. Kovaleva, et al., *Optical response of ferromagnetic YTiO$_3$ studied by spectral ellipsometry.* Physical Review B, 2007. **76**(15).
23. C. Ulrich, et al., *Raman Scattering in the Mott Insulators LaTiO$_3$ and YTiO$_3$: Evidence for Orbital Excitations.* Physical Review Letters, 2006. **97**(15).
24. T. Kiyama, et al., *Orbital Fluctuations in Ground State of YTiO$_3$: $^{47,49}$Ti NMR Study.* Journal of the Physical Society of Japan, 2005. **74**(4): pp. 1123-1126.
25. M. Tsubota, et al., *Low-field magnetic anisotropy in Mott-insulating ferromagnet $Y_{1-x}Ca_xTiO_3$ (x⩽0.1).* Physica B: Condensed Matter, 2000. **281-282**: pp. 622-624.
26. W. Knafo, et al., *Ferromagnetism and lattice distortions in the perovskite YTiO$_3$.* Physical Review B, 2009. **79**(5).
27. P. Yordanov. *Spectroscopic Study of CaMnO$_3$/CaRuO$_3$ Superlattices and YTiO3 Single Crystals* Ph.D. thesis, Max Planck Institute for Solid State Research, (2009).
28. M. Först, et al., *Nonlinear phononics as an ultrafast route to lattice control.* Nature Physics, 2011. **7**(11): pp. 854-856.
29. A. Subedi, A. Cavalleri, and A. Georges, *Theory of nonlinear phononics for coherent light control of solids.* Physical Review B, 2014. **89**(22).
30. P.G. Radaelli, *Breaking symmetry with light: Ultrafast ferroelectricity and magnetism from three-phonon coupling.* Physical Review B, 2018. **97**(8).
31. R. Mankowsky, et al., *Nonlinear lattice dynamics as a basis for enhanced superconductivity in YBa2Cu3O6.5.* Nature, 2014. **516**(7529): pp. 71-73.
32. M. Rini, et al., *Control of the electronic phase of a manganite by mode-selective vibrational excitation.* Nature, 2007. **449**(7158): pp. 72-74.
33. M. Först, et al., *Driving magnetic order in a manganite by ultrafast lattice excitation.* Physical Review B, 2011. **84**(24).
34. A.S. Disa, et al., *Polarizing an antiferromagnet by optical engineering of the crystal field.* Nature Physics, 2020. **16**(9): pp. 937-941.
35. D. Afanasiev, et al., *Ultrafast control of magnetic interactions via light-driven phonons.* Nature Materials, 2021. **20**(5): pp. 607-611.
36. G. Khalsa and N.A. Benedek, *Ultrafast optically induced ferromagnetic/anti-ferromagnetic phase transition in GdTiO3 from first principles.* npj Quantum Materials, 2018. **3**(1).
37. M. Gu and J.M. Rondinelli, *Nonlinear phononic control and emergent magnetism in Mott insulating titanates.* Physical Review B, 2018. **98**(2).
38. N.N. Kovaleva, et al., *Dipole-active optical phonons in YTiO$_3$: Ellipsometry study and lattice-dynamics calculations.* Physical Review B, 2009. **79**(4).
39. B. Liu, et al., *Generation of narrowband, high-intensity, carrier-envelope phase-stable pulses tunable between 4 and 18 THz.* Optics Letters, 2016. **42**(1).
40. The bandwidth of 17 THz excitation spans both of the highest modes in YTiO3. They are found to have similar effects on the magnetic properties, and our theoretical analysis takes into account the contributions of both modes.
41. K.I. Kugel and D.I. Khomskii, *Crystal structure and magnetic properties of substances with orbital degeneracy.* Zh. Eksp. Teor. Fiz., 1972. **64**: pp. 1429-1439.
42. M. Matsubara, et al., *Ultrafast Photoinduced Insulator-Ferromagnet Transition in the Perovskite ManganiteGd0.55Sr0.45MnO3.* Physical Review Letters, 2007. **99**(20).
43. S.F. Maehrlein, et al., *Dissecting spin-phonon equilibration in ferrimagnetic insulators by ultrafast lattice excitation.* Science Advances, 2018. **4**(7).
44. Z. Sun and A.J. Millis, *Pump-induced motion of an interface between competing orders.* Physical Review B, 2020. **101**(22).
45. Z. Sun and A.J. Millis, *Transient Trapping into Metastable States in Systems with Competing Orders.* Physical Review X, 2020. **10**(2).

46. D.M. Juraschek, Q.N. Meier, and P. Narang, *Parametric Excitation of an Optically Silent Goldstone-Like Phonon Mode.* Physical Review Letters, 2020. **124**(11).
47. R.M. Fernandes, A.V. Chubukov, and J. Schmalian, *What drives nematic order in iron-based superconductors?* Nature Physics, 2014. **10**(2): pp. 97-104.
48. K. Jin, et al., *Link between spin fluctuations and electron pairing in copper oxide superconductors.* Nature, 2011. **476**(7358): pp. 73-75.
49. T. Moriya and K. Ueda, *Spin fluctuations and high temperature superconductivity.* Advances in Physics, 2000. **49**(5): pp. 555-606.
50. R. Arpaia, et al., *Dynamical charge density fluctuations pervading the phase diagram of a Cu-based high-Tc superconductor.* Science, 2019. **365**(6456): pp. 906-910.
51. M. Mitrano, et al., *Possible light-induced superconductivity in K3C60 at high temperature.* Nature, 2016. **530**(7591): pp. 461-464.
52. T.F. Nova, et al., *Metastable ferroelectricity in optically strained SrTiO3.* Science, 2019. **364**(6445): pp. 1075-1079.
53. A. Kogar, et al., *Light-induced charge density wave in LaTe$_3$.* Nature Physics, 2019. **16**(2): pp. 159-163.